\documentclass[aps,prb,twocolumn,showpacs,superscriptaddress]{revtex4-1}
\usepackage{amssymb,amsmath}
\usepackage{graphicx}
\usepackage{dcolumn}
\usepackage{bm}
\usepackage{color}

\begin{document}

\title{First-principles modeling of three-body interactions in highly compressed solid helium}

\author{Claudio Cazorla}
\email{c.cazorla@unsw.edu.au}
\affiliation{School of Materials Science and Engineering, UNSW Australia, Sydney NSW 2052, Australia \\
Integrated Materials Design Centre, UNSW Australia, Sydney NSW 2052, Australia}

\author{Jordi Boronat}
\email{jordi.boronat@upc.edu}
\affiliation{Departament de F\'{i}sica i Enginyeria Nuclear, Universitat Polit\`{e}cnica
de Catalunya, Campus Nord B4-B5, E-08034, Barcelona, Spain}

\email{c.cazorla@unsw.edu.au}

\begin{abstract}
We present a new set of three-body interaction models based on the Bruch-McGee (BM) 
potential that are suitable for the study of the energy, structural and elastic properties
of solid $^{4}$He at high pressure. Our \emph{ab initio} three-body potentials are obtained from 
the fit to total energies and atomic forces computed with the van der Waals density 
functional theory method due to Grimme, and represent an improvement with respect to 
previously reported three-body interaction models. In particular, we show that some of the introduced 
BM parametrizations reproduce closely the experimental equation of state and bulk modulus 
of solid helium up to a pressure of $\sim 60$~GPa, when used in combination with standard pairwise 
interaction models in diffusion Monte Carlo simulations. Importantly, we find that recent 
predictions reporting a surprisingly small variation of the kinetic energy and
Lindeman ratio on quantum crystals under increasing pressure are likely to be artifacts produced 
by the use of incomplete interaction models. Also, we show that the experimental variation of the 
shear modulus, $C_{44}$, at $P \le 25$~GPa can be quantitatively described with the new 
set of three-body BM potentials. At higher pressures, however, the agreement between our $C_{44}$ 
results and experiments deteriorates and thus we argue that higher order many-body terms in the expansion 
of the atomic interactions probably are necessary in order to better describe elasticity in very dense 
solid $^{4}$He. 
\end{abstract}

\pacs{67.80.-s,02.70.Ss,67.40.-w}
\maketitle

\section{Introduction}
\label{sec:introduction}
The electronic structure of a single $^{4}$He atom is among the simplest in 
the periodic table of elements. Likewise, the atomic interactions in liquid 
and solid helium can be reproduced accurately with simple analytical 
functions that solely depend on the distance between particles taken in pairs. 
Examples of successful $^{4}$He--$^{4}$He interaction models include the 
Lennard-Jones and Aziz-type semiempirical potentials.~\cite{kalos81,boronat94,aziz}  
Yet, under conditions of large pressures and strain deformations the interparticle 
interactions become more complex due to the strong electronic repulsion experienced 
by neighboring atoms. Consequently, pairwise potentials, which work reasonably well
under near-equilibrium conditions, turn out to be unreliable.
This is, for instance, the case of the Aziz-II potential,~\cite{aziz} which 
at high pressure provides too repulsive atomic forces and a significant 
overestimation of the $^{4}$He molar volume and bulk modulus.~\cite{cazorla08} 

A recently proposed straightforward way to correct for such modeling drawbacks consists in 
modifying the repulsive part of standard pairwise potentials by means of an exponential 
attenuation factor.~\cite{moraldi12} This possibility has already been explored in highly 
compressed solid $^{4}$He~\cite{azizbc} and molecular hydrogen~\cite{omiyinka13} with quantum 
Monte Carlo simulations, producing equations of state which are in very good agreement with 
experiments. Nevertheless, the use of modified pairwise potentials in very dense crystals 
poses a series of issues and open questions. For instance, a surprisingly small variation of 
the kinetic energy upon increasing pressure have been reported in works~[\onlinecite{azizbc}]
and~[\onlinecite{omiyinka13}], and, owing to the lack of experimental data in the thermodynamic 
regime of interest, it remains to be demonstrated whether such predictions can be fully ascribed to 
genuine quantum nuclear effects or not. Also, pairwise potentials are in general not recommended 
for the study of elasticity in hcp crystals at high pressure since they inevitably lead to null values 
of the Cauchy discrepancy (defined as the difference between the two elastic constants $C_{12}$ 
and $C_{44}$), in contrast to what is observed in experiments.~\cite{wallace72,pechenik08,freiman09,zha04} 

An alternative route to improve the description of quantum solids under extreme stress-strain 
conditions is to consider higher order terms, beyond pairwise additivity, in the 
approximation to the atomic interactions. In this context, several three-body interatomic 
models have already been proposed like, for instance, the Axilrod-Teller 
(AT), Bruch-McGee (BM), and Cohen-Murrel (CM) potentials.~\cite{boronat94,bm73,cohen96} 
However, improvements resulting from the use of those three-body interaction models so
far have been reported to be only marginal. For instance, three decades ago Loubeyre claimed, based on the 
outcomes of self-consistent phonon and classical Monte Carlo simulations, that the BM three-body 
interaction could bring into good agreement calculations and experiments performed on the 
equation of state of solid helium up to $\sim 60$~GPa.~\cite{loubeyre86} 
However, Chang \emph{et al.}~\cite{chang01} have shown more recently that when either the BM or CM 
three-body potentials are considered in quantum Monte Carlo simulations the resulting $^{4}$He 
molar volumes are significantly underestimated, already at few GPa. Similar discouraging results 
have been reported also by other authors who have employed analogous three-body interaction models.~\cite{herrero06,tian06}    

In this article, we present new work done on the modeling of three-body interactions in highly 
compressed solid helium up to pressures of $\sim 160$~GPa. We introduce a new set of BM 
potential parametrizations obtained from fits to \emph{ab initio} energies and atomic forces 
calculated with the van der Waals corrected density functional theory method due to Grimme
(DFT-D2).~\cite{grimme06} We show that an overall improved description of the energy, 
elastic and structural properties of solid helium can be achieved with some of the 
introduced BM three-body interatomic potentials, when used in combination with pairwise 
potentials in quantum Monte Carlo simulations. Our work also brings new insight 
into the physics of quantum crystals at high pressure. For instance, we show that previously 
reported small variations of the kinetic energy, ${\rm E_{k}}$, and Lindeman ratio, $\gamma$, 
in solid helium under pressure~\cite{azizbc} are likely to be artifacts deriving 
from the use of incomplete atomic interaction models. Moreover, we quantify the role of quantum 
nuclear effects on the estimation of the shear modulus, $C_{44}$, and conclude that they become 
secondary when pressure is raised. Finally, at $P \sim 25$~GPa we find 
that the agreement between our $C_{44}$ results and experiments starts 
to worsen. Therefore, we argue that higher order many-body terms in the expansion of the atomic 
interactions probably are necessary in order to describe elasticity in dense solid helium more 
accurately.   

The organization of this article is as follows. In the next section, we outline the 
employed computational methods and provide the technical details in our calculations. 
In Sec.~\ref{sec:potentialfit}, we explain the fitting strategy that we have 
followed to obtain the new set of three-body interaction models. 
Next, we present our results on the equation of state, kinetic energy, and 
structural and elastic properties in solid helium, together with some discussion. 
Finally, we summarize our main findings in Sec.~\ref{sec:conclusions}.

\section{Computational Methods}
\label{sec:methods}
We used the density functional theory method including van der Waals corrections due
to Grimme,~\cite{grimme06} to compute the interactions and forces between helium atoms in the
hexagonal close package (hcp) crystal structure, from equilibrium up to a pressure of
$\sim 160$~GPa. (Details of our ab initio DFT-D2 calculations can be found in elsewhere,~\cite{azizbc}
hence we highlight here only the main technical features.)
Subsequently, we found a series of three-body interaction models that, when used in combination with 
the pairwise Aziz-II potential~\cite{aziz} (hereafter denoted as $V_{2}$), reproduced very closely 
the obtained DFT-D2 results. The details of our fitting strategy are comprehensively explained in 
Sec.~\ref{sec:potentialfit}. Next, we performed diffusion Monte Carlo (DMC) simulations in which
the new three-body interaction models were used to calculate the energy, structural, 
and elastic properties of solid helium under pressure. In this section, we explain the specific
implementation of the DFT-D2 and DMC methods in our work.

\subsection{Density functional theory}
\label{subsec:dft}
We chose the generalized gradient approximation to density functional theory proposed by 
Perdew, Burke, and Ernzerhof (GGA-PBE),~\cite{pbe96} as is implemented in the VASP package.~\cite{vasp}
Van der Waals interactions were taken into account by adding an attractive energy term to the 
exchange-correlation energy of the form $E_{\rm disp} = -\sum_{i,j} C_{6} / r^{6}_{ij}$ (where 
indexes $i$ and $j$ label different particles, $C_{6}$ is a constant, and a damping factor is 
introduced at short distances to avoid divergences).~\cite{grimme06,cazorla15} 
The projector-augmented-wave technique~\cite{blochl94,kresse99} was employed to represent 
the core electrons since this approach has been shown to provide very accurate total energies 
and is computationally very efficient.~\cite{cazorla07,taioli07}
The electronic wave functions were represented in a plane-wave basis truncated at $500$~eV, 
and for integrations within the first Brillouin zone (BZ) we employed dense $\Gamma$-centered 
$k$-point grids of $14 \times 14 \times 14$. By using these parameters we obtained interaction
energies that were converged to within $5$~K per atom. Geometry relaxations were performed by 
using a conjugate-gradient algorithm that kept the volume of the unit cell fixed and permitted 
variations of its shape. The imposed tolerance on the atomic forces was $0.005$~eV$\cdot$\AA$^{-1}$. 
With such a DFT-D2 setup we calculated the total energy and shear modulus in solid $^{4}$He in the 
volume interval $3 \le V \le 16$~\AA$^{3}$/atom. 

Additionally, we computed the vibrational phonon spectrum in solid $^{4}$He at eight different volumes
by means of the ``direct approach''. In the direct approach the force-constant matrix is directly calculated 
in real-space by considering the proportionality between atomic displacements and forces when the 
former are sufficiently small.~\cite{cazorla13b,shevlin12,cazorla08c} In this case, large supercells have
to be simulated in order to guarantee that the elements of the force-constant matrix have all fallen 
off to negligible values at their boundaries, a condition that follows from the use of periodic boundary 
conditions.~\cite{alfe09} Once the force-constant matrix is obtained, we Fourier-transform it
to obtain the phonon spectrum at any $q$-point. The quantities with respect to which our DFT-D2 phonon  
calculations need to be converged are the size of the supercell and atomic displacements,
and the numerical accuracy in the atomic forces. The following settings were found to fulfill our 
convergence requirement of correct zero-point energy corrections to within $5$~K/atom:~\cite{azizbc,cazorla13b} 
$4 \times 4 \times 3$ supercells (that is, $48$ repetitions of the hcp unit cell containing a total 
of $96$ atoms), and atomic displacements of $0.02$~\AA~. Regarding the calculation of the atomic forces 
with VASP, we found that the density of $k$-points had to be increased slightly with respect to the value 
used in the energy calculations (i.e., from $14 \times 14 \times 14$ to $16 \times 16 \times 16$) and that 
computation of the non-local parts of the pseudopotential contributions needed to be performed in reciprocal, 
rather than real, space.

\subsection{Diffusion Monte Carlo}
\label{subsec:dmc}
In our DMC simulations, we used a guiding wave function, $\Psi_{\rm SNJ}$, that accounts simultaneously
for the atomic periodicity and Bose-Einstein quantum symmetry in $^{4}$He crystals. This model wave 
function is expressed as~\cite{cazorla09b}  
\begin{equation}
\Psi_{\rm SNJ}({\bf r}_1,\ldots,{\bf r}_N) = \prod_{i<j}^{N} f(r_{ij}) 
\prod_{J=1}^{N} \left( \sum_{i=1}^{N} g(r_{iJ}) \right)~,
\label{snjtrial}
\end{equation}
where indexes $\lbrace i,j \rbrace$ and $J$ run over particles and perfect lattice positions, 
respectively. In previous works we have shown that $\Psi_{\rm SNJ}$ provides an excellent 
description of the ground-state properties of bulk hcp $^{4}$He and other similar quantum 
systems.~\cite{cazorla09b,cazorla08b,cazorla10,gordillo11}
The correlation factors in Eq.~(\ref{snjtrial}) were expressed in the McMillan, 
$f(r) = \exp\left[-1/2~(b/r)^{5}\right]$~, and Gaussian, $g(r) = \exp\left[-1/2~(a r^{2})\right]$, 
forms. Parameters $a$ and $b$ were optimized at each density point by using the variational 
Monte Carlo (VMC) method. For instance, at $\rho = 0.06$~\AA$^{-3}$~ we obtained $b = 2.94$~\AA~ 
and $a = 3.21$~\AA$^{-2}$~, and at $\rho = 0.33$~\AA$^{-3}$~, $b = 1.84$~\AA~ 
and $a = 29.08$~\AA$^{-2}$~. 
We note that our choice of the guiding function was motivated by an interest in studying the 
possible effects of quantum atomic exchanges on the energetic and elastic properties of dense helium. 
However, we realised by direct comparison to the results obtained with non-symmetric wave function models 
in analogous DMC simulations,~\cite{azizbc} that such effects can be totally neglected in practice. 

The technical parameters in our calculations were set to ensure convergence of the total 
energy per particle to less than $5$~K. The value of the mean population of walkers was $10^{3}$ 
and the length of the imaginary time-step ($\Delta \tau$) $10^{-4}$~K$^{-1}$~. We used simulation 
cells containing $180$ atoms. Numerical bias stemming from the finite size of the simulation box 
were minimised by following the variational correction approach explained in works~[\onlinecite{cazorla08}]
and~[\onlinecite{azizbc}]. 
Statistics were accumulated over $10^{5}$ DMC steps performed after system equilibration, and the 
approximation used for the short-time Green's function, $e^{-\hat{H} \tau}$, was accurate to second order 
in $\Delta \tau$.~\cite{boronat94,chin90} The computational strategy that we followed to calculate the shear 
modulus $C_{44}$ was the same than in Refs.~[\onlinecite{cazorla12,cazorla13,cazorla12b}]. 

\begin{figure}
\centerline
        {\includegraphics[width=1.0\linewidth]{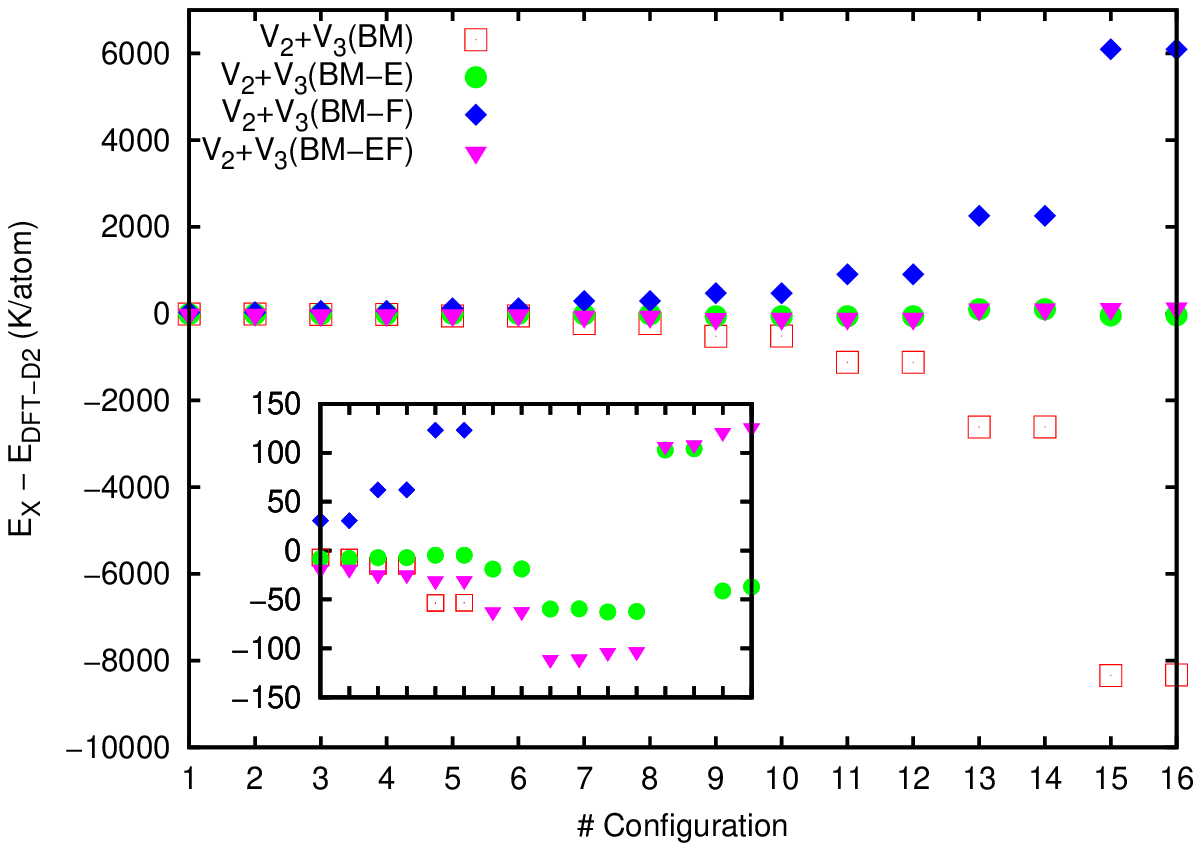}}
        {\includegraphics[width=1.0\linewidth]{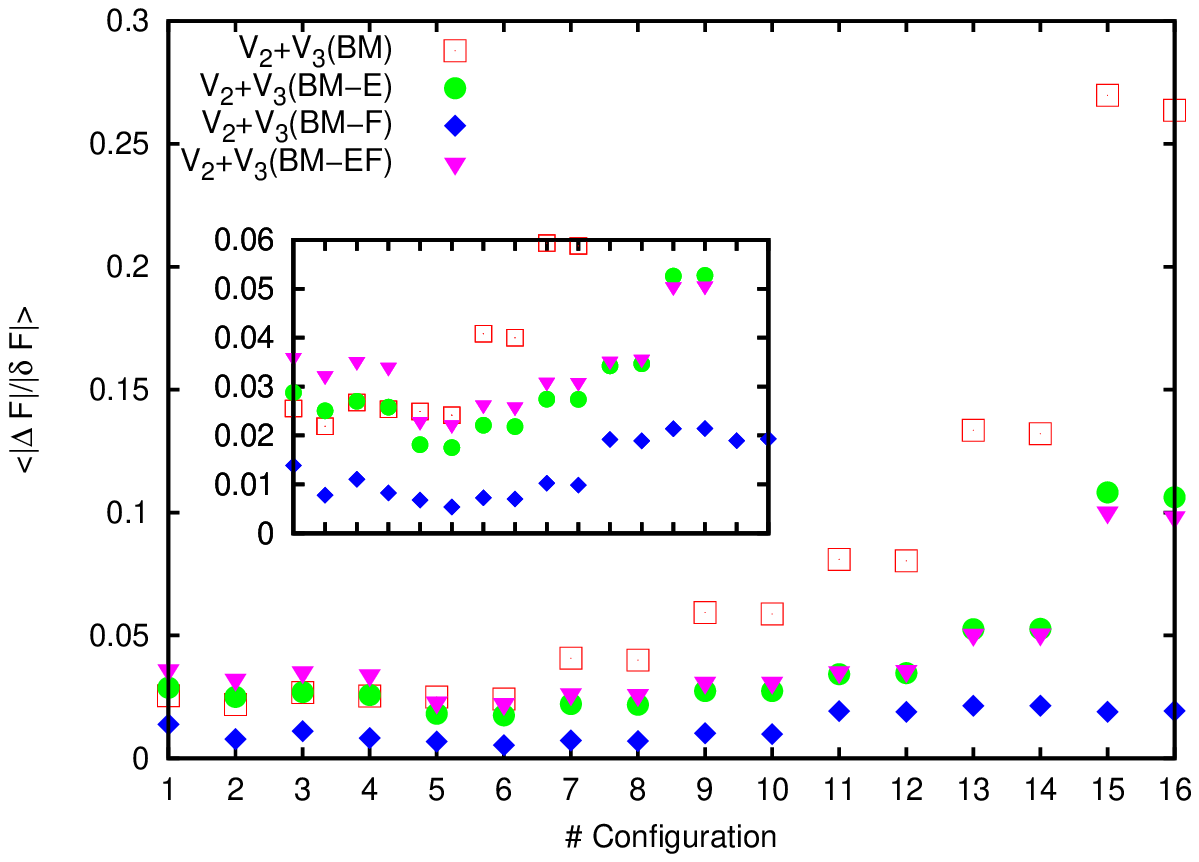}}
\caption{(\emph{Top})~Energy differences between the DFT-D2 method and $V_{3}$ potentials 
calculated on a reference set of $16$ configurations (see text). Details are magnified 
in the inset. (\emph{Bottom})~Results of our fit obtained in the case of the atomic forces. 
$ \Delta F$ stands for the difference in the atomic forces between the DFT-D2 method and 
many-body potentials, $\delta F$ for the variance of the atomic forces computed with the DFT-D2 
method, and $\langle \cdots \rangle$ for the average performed over particles and Cartesian components.}
\label{potfit}
\end{figure}

\begin{figure}
\centerline
        {\includegraphics[width=1.0\linewidth]{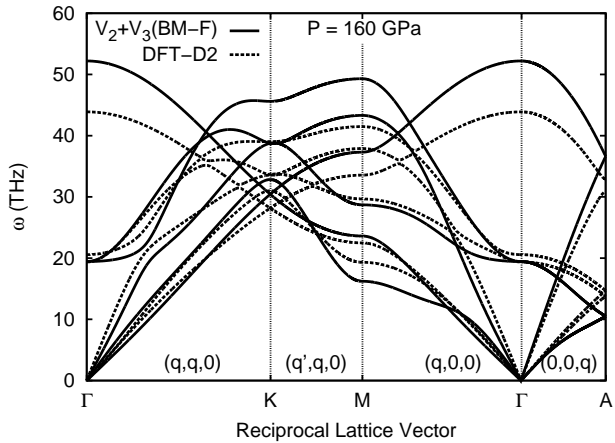}}
\caption{Phonon spectrum obtained with the DFT-D2 method (dashed lines) and the $V_{2} + V_{3}$(BM-F)
interaction model (solid lines), which was determined considering only the atomic forces in the
corresponding fit (see text).}
\label{phonons}
\end{figure}

\section{Fitting strategy and three-body potential models}
\label{sec:potentialfit}
Our three-body potential matching algorithm~\cite{ercolessi94,marco12,marco14} is based on a least square fit 
to the DFT-D2 reference data, that consists of total energies and atomic forces. The objective function to be 
minimized is given by  
\begin{eqnarray}
&& \chi^{2} = \omega_{E} \times \sum_{i}^{N}  \frac{ \left( E_{i}^{{\rm FF}} - E_{i}^{{\rm DFT}} \right)^{2}}{\sum_{j}^{N}  \left( E_{j}^{{\rm DFT}} - \langle E^{{\rm DFT}} \rangle \right)^{2}}   \nonumber \\
&& + \quad \omega_{F} \times \sum_{i}^{N} \frac{\sum_{l,\alpha}^{n,3} \left( F_{l \alpha,i}^{{\rm FF}} - F_{l \alpha,i}^{{\rm DFT}}\right)^{2}}{\sum_{l,\alpha,j}^{n,3,N} \left( F_{l \alpha,j}^{{\rm DFT}} - \langle F^{{\rm DFT}} \rangle   \right)^{2}}~, 
\label{eq:fitfunct}
\end{eqnarray}
where $N=16$ is the number of reference configurations, $n = 96$ the number of particles on  
each configuration, and $\omega_{E}$ and $\omega_{F}$ a weight assigned to the energy, $E$, and force, 
$F$, contributions to $\chi^{2}$, respectively. With this definition of the objective function we ensure that 
despite different magnitudes are expressed in different units all them are normalized and contribute equally 
to $\chi^{2}$. Subscripts ``DFT'' and ``FF'' refer to the DFT-D2 and classical potential results, respectively.

The set of reference configurations in our fit comprised the $16$ structures used in the calculation 
of the $^{4}$He vibrational phonon spectra in the interval $3 \le V \le 16$~\AA$^{3}$/atom by 
means of the ``direct approach'' (see Sec.~\ref{subsec:dft}).~\cite{cazorla13b,shevlin12,cazorla08c}
Such atomic arrangements were generated by taking the relaxed hcp lattice supercells ($P6_{3}/mmc$, 
space group $194$) at $8$ different volumes and displacing one of the atoms sitting in an inequivalent 
$d$ Wyckoff position a distance of $0.02$~\AA~ first along the $\frac{1}{2} {\bf \hat{x}} - \frac{\sqrt{3}}{2} {\bf \hat{y}}$ 
direction (where ${\bf \hat{x}}, {\bf \hat{y}}, {\bf \hat{z}}$ represent the normalised Cartesian vectors),  
and then along ${\bf \hat{z}}$ (that is, we created two different atomic configurations at each volume).   
The reason for our choice was that we wanted to reproduce simultaneously the energy and elastic properties 
in highly compressed solid $^{4}$He. In fact, the atomic forces are defined as minus the first derivative of the 
total energy with respect to the atomic positions, whereas the elastic constants involve the second derivative 
of the total energy with respect to strain deformations. In spite of this apparent disconnection, atomic forces 
and elastic constants are indirectly related by the corresponding spectrum of vibrational phonon frequencies. 
Namely, on one side, phonons can be calculated from the variation of the atomic forces upon the displacement of atoms 
away from their equilibrium positions, and, on the other side, elastic constants can be estimated from the slope of 
specific acoustic branches in the vicinity of the $\Gamma$ point in reciprocal space (that is, in the $q \to 0$ limit). 
Therefore, even though we did not explicitly consider second derivatives in our definition of the objective function $\chi^{2}$, 
we expected to achieve an acceptable description of elasticity in solid helium. We shall come back to this point 
later on this section.        

The classical potential adopted in this study, denoted as ``FF'' in Eq.~\ref{eq:fitfunct}, is given by 
$U_{\rm pot} = V_{2} + V_{3}$, where $V_{2}$ represents the pairwise Aziz-II interaction model~\cite{aziz} 
and $V_{3}$ the three-body Bruch-McGee (BM) potential given by~\cite{bm73} 
\begin{eqnarray} 
V_{3} \left( r_{ij}, r_{ik}, r_{jk} \right) &&= \Big[ \frac{\nu}{r_{ij}^{3} r_{ik}^{3} r_{jk}^{3}} - A \exp{\left( -\alpha [ r_{ij} + r_{ik} + r_{jk} ] \right) } \Big]  \nonumber \\
&& \times \left( 1 + 3 \cos{\phi_{i}} \cos{\phi_{j}} \cos{\phi_{k}}  \right)~,  
\label{eq:bmpot}
\end{eqnarray}
where $r_{ij} = \mid {\bf r}_{i} - {\bf r}_{j} \mid $, and $\phi_{i}$, $\phi_{j}$, and $\phi_{k}$,
are the interior angles of the triangle formed by the atoms labelled $i$, $j$, and $k$.
$V_{3}$ is an attractive potential term representing triple dipole and three-body exchange interactions.  
Parameters $\nu$, $A$, and $\alpha$ were varied during the minimization 
of the objective function $\chi^{2}$ (see Eq.~\ref{eq:fitfunct}). For this, we used a quadratic 
polynomial interpolation line-search with the directions found using the Broyden-Fletcher-Goldfarb-Shanno 
(BFGS) formula.~\cite{nocedal06} The gradient of the objective function was calculated analytically since 
otherwise numerical bias developed that impeded convergence. Actually, the typical size of the involved 
atomic forces is very small, of the order of $0.01-0.1$~eV/\AA~, hence they needed to be calculated very precisely. 
The minimizations were stopped when all the gradients of the objective function in absolute value were smaller 
than $10^{-5}$. Typically, this was achieved within $\sim 100$ minimization loops when starting from a reasonable initial 
guess of the $\nu$, $A$, and $\alpha$ parameters (e.g., the original values proposed by Bruch and 
McGee~[\onlinecite{bm73}]).     

\begin{table*}
\begin{center}
\label{tab:bmparameters}
\begin{tabular}{c c c c}
\hline
\hline
$ $ & $ $ & $ $ & $ $ \\
$\qquad \quad$ & $\qquad \nu$~(K$\cdot \sigma^{9}$) \quad & $\qquad A$~(K) \quad & $\qquad \alpha$~($\sigma^{-1}$) \qquad \\
$ $ & $ $ & $ $ & $ $ \\
\hline
$ $ & $ $ & $ $ & $ $ \\
$V_{3}$(BM)~[\onlinecite{bm73}] & $0.3270  $ & $9~676~545.53  $ & $4.9480 $ \\
$ $ & $ $ & $ $ & $ $ \\
$V_{3}$(BM-E)       & $-0.4910 $ & $14~754~161.38 $ & $5.6128 $ \\
$ $ & $ $ & $ $ & $ $ \\
$V_{3}$(BM-F)       & $1.4029  $ & $12~863~029.73 $ & $5.8273 $ \\
$ $ & $ $ & $ $ & $ $ \\
$V_{3}$(BM-EF)     & $-1.1364 $ & $29~189~436.37 $ & $6.0691 $ \\
$ $ & $ $ & $ $ & $ $ \\
\hline
\hline
\end{tabular}
\end{center}
\caption{Bruch-McGee three-body potential parameters corresponding to the original
model, $V_{3}$(BM), and the new ones introduced in the present work. The $V_{3}$(BM-E) set has
been obtained by considering exclusively DFT-D2 energies on the fit [$\omega_{E} = 1$, 
$\omega_{F} = 0$], the $V_{3}$(BM-F) the atomic forces [$\omega_{E} = 0$,
$\omega_{F} = 1$], and $V_{3}$(BM-EF) a combination of \emph{ab initio} energies 
and atomic forces [$\omega_{E} = 0.5$, $\omega_{F} = 0.5$] (see text).}
\end{table*}

Table~I shows the values of the parameters obtained in our $V_{3}$ fits, in which 
we considered three different possibilities based on the choice of the relative 
energy and forces weights:~$(1)$~$\omega_{E} = 1$ and $\omega_{F} = 0$,
hereafter denoted as $V_{3}$(BM-E), $(2)$~$\omega_{E} = 0$ and $\omega_{F} = 1$,
$V_{3}$(BM-F), and $(3)$~$\omega_{E} = 0.5$ and $\omega_{F} = 0.5$, $V_{3}$(BM-EF). 
Our results differ appreciably from the original values 
proposed by Brunch and McGee [which hereafter are denoted as $V_{3}$(BM)]. For 
instance, $\nu$ becomes negative when the energies are taken into account 
in the fit, and $A$ and $\alpha$ systematically turn out to be larger. 

In Figure~\ref{potfit}, we illustrate the quality of our fits by plotting the energies
and forces calculated on each reference configuration. For comparison purposes, we also enclose the  
results obtained with the original $V_{3}$(BM) potential (i.e., with function $U_{\rm pot} = V_{2} + V_{3}$). 
For the sake of simplifying the notation, we only indicate the three-body part in the 
corresponding many-body potential. This convention will be adopted throughout the text if not 
stated otherwise. As is appreciated in the figure, $V_{3}$(BM-E) reproduces the DFT-D2 energies more 
closely than any other model (as expected) whereas $V_{3}$(BM) provides the worst description. 
The energies obtained with the $V_{3}$(BM-EF) potential can be regarded also as fairly 
good. As for the atomic forces, $V_{3}$(BM-F) produces the best results, as expected,  
and $V_{3}$(BM), again, turns out to be the less reliable. In this latter case, the forces 
obtained with the $V_{3}$(BM-EF) and, surprisingly also, $V_{3}$(BM-E) potentials are not 
too distant from the reference DFT-D2 data.        

In Fig~\ref{phonons}, we plot the vibrational phonon spectra obtained with the DFT-D2 method
and the $V_{3}$(BM-F) potential in solid $^{4}$He at the highest analysed pressure (probably the 
most challenging case to be reproduced with a potential function, see Fig.~\ref{potfit}).  
We note that the agreement between the two sets of data can be regarded as fairly good. 
The largest differences are found on the optical branches, which correspond to the highest vibrational 
frequency values. The DFT-D2 acoustic phonon modes in the vicinity of the $\Gamma$ point, however, are 
reasonably well reproduced by $V_{3}$(BM-F). These outcomes demonstrate that, as we suggested above, 
by considering the atomic forces in the definition of $\chi^{2}$ in principle one can obtain a 
reasonable description of the elasticity in the reference system.        

\begin{figure}
\centerline
        {\includegraphics[width=1.0\linewidth]{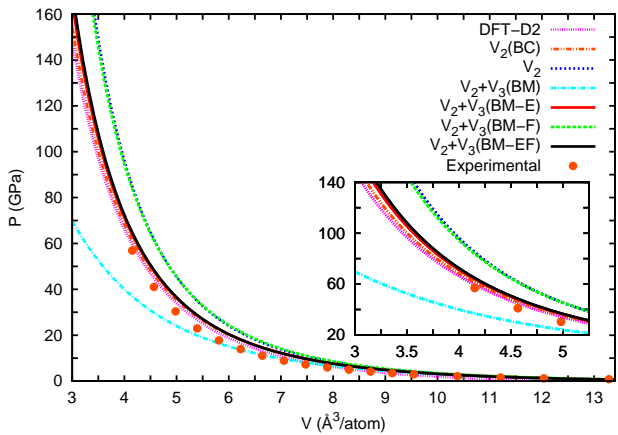}}
\caption{Zero-temperature equation of state calculated in helium with the DFT-D2 and DMC 
         methods. In the DMC case, different pairwise and three-body interaction models
         have been employed. Experimental data from Ref.~[\onlinecite{loubeyre93}]
         are shown for comparison. \emph{Inset}: The high-$P$ region in the $P(V)$ curves 
         are magnified in order to appreciate better the differences.}
\label{eos}
\end{figure}

\section{Results and Discussion}
\label{sec:results}

\subsection{Equation of state}
\label{subsec:energy}
We show the results of our calculations on the equation of state, $P(V)$, of solid helium in 
Fig.~\ref{eos}, together with experimental data from work~[\onlinecite{loubeyre93}].
The DFT-D2 series was obtained with the \emph{ab initio} methods explained in Sec.~\ref{subsec:dft}, 
including quantum zero-point energy corrections. The other results were obtained 
in diffusion Monte Carlo (DMC) simulations using the indicated interaction potentials, 
as explained in Sec.~\ref{subsec:dmc} and elsewhere~[\onlinecite{azizbc}]. Labels ``$V_{2}$'' and 
``$V_{2}$(BC)'' stand respectively for the pairwise potential due to Aziz~\cite{aziz} and a modified 
version of the former that we have recently introduced in work~[\onlinecite{azizbc}]. The DMC (DFT-D2) calculations 
were performed at $12$~($8$) different volumes spanned in the interval $3 \le V \le 16$~\AA$^{3}$/atom. 
In each case, the resulting total energies were fitted to a third order 
Birch-Murnaghan equation of the form~\cite{birch78,cazorla09}
\begin{eqnarray} 
&&E(V) - E_{0} = \frac{3}{2}~V_{0}~B_{0} \times \nonumber \\  
&& \bigg [ -\frac{\chi}{2} \left ( \frac{V_{0}}{V} \right )^2 + \frac{3}{4}~ \left ( 1+2 \chi \right ) \left ( \frac{V_{0}}{V} \right )^{(4/3)} \nonumber \\
&& - \frac{3}{2} \left ( 1+\chi \right ) \left (\frac{V_{0}}{V} \right )^{(2/3)} + \frac{1}{2} \left (\chi+\frac{3}{2}\right ) \bigg ]~,
\label{eq:eqstate}
\end{eqnarray}
where $B_{0}= V_{0}\frac{d^2E}{dV^2}$ is the value of the bulk modulus at the equilibrium volume $V_{0}$,  
$\chi = \frac{3}{4}\left ( 4 - B^{'}_{0} \right )$ with $B^{'}_{0}=\left(d B_{0}/d P\right)$, 
and all the derivatives are calculated at zero pressure. For reproducibility purposes, we enclose the 
$V_{0}$, $B_{0}$, and $B^{'}_{0}$ parameters obtained in all our fits in Table~II.

\begin{table*}
\begin{center}
\label{tab:eosfit}
\begin{tabular}{c c c c}
\hline
\hline
$ $ & $ $ & $ $ & $ $ \\
$\qquad \quad$ & $\qquad V_{0}$~(\AA$^{3}$) \quad & $\qquad B_{0}$~(eV/\AA$^{3}$) \quad & $\qquad B^{'}_{0}$ \qquad \\
$ $ & $ $ & $ $ & $ $ \\
\hline
$ $ & $ $ & $ $ & $ $ \\
${\rm DFT-D2}$          & $ 12.23 $ & $ 0.0398 $ & $ 3.9648 $ \\
$ $ & $ $ & $ $ & $ $ \\
$V_{2}$(BC)             & $ 15.68 $ & $ 0.0166 $ & $ 4.1144 $ \\
$ $ & $ $ & $ $ & $ $ \\
$V_{2}$                 & $ 16.61 $ & $ 0.0115 $ & $ 4.8829 $ \\
$ $ & $ $ & $ $ & $ $ \\
$V_{2}+V_{3}$(BM)       & $ 15.68 $ & $ 0.0181 $ & $ 3.6722 $ \\
$ $ & $ $ & $ $ & $ $ \\
$V_{2}+V_{3}$(BM-E)     & $ 15.84 $ & $ 0.0165 $ & $ 4.1854 $ \\
$ $ & $ $ & $ $ & $ $ \\
$V_{2}+V_{3}$(BM-F)     & $ 16.58 $ & $ 0.0130 $ & $ 4.6709 $ \\
$ $ & $ $ & $ $ & $ $ \\
$V_{2}+V_{3}$(BM-EF)    & $ 15.85 $ & $ 0.0158 $ & $ 4.2463 $ \\
$ $ & $ $ & $ $ & $ $ \\
\hline
\hline
\end{tabular}
\end{center}
\caption{Parameters corresponding to the fit of our equation of state results to Birch-Murnaghan 
functions, see Eq.~(\ref{eq:eqstate}), as obtained with different computational approaches. 
In the DMC case, different pairwise and three-body potentials have been considered for the
description of the interatomic forces.}
\end{table*}

\begin{figure}
\centerline
        {\includegraphics[width=1.0\linewidth]{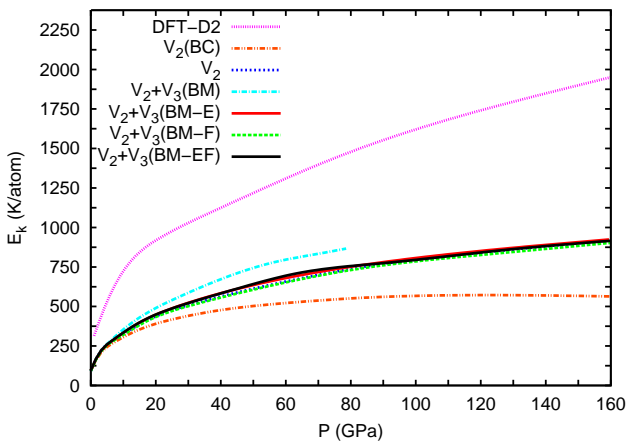}}
\caption{Atomic kinetic energy calculated in $^{4}$He with the DFT-D2 and DMC
         methods and expressed as a function of pressure. In the DMC case,
         different pairwise and three-body interaction models have been considered
         for the description of the atomic interactions.}
\label{kinetice}
\end{figure}

Very good agreement is obtained between our DFT-D2 results and experiments. This outcome
justifies in part our choice of the DFT-D2 results as reference data in modeling of the many-body 
interactions. Likewise, the $P(V)$ curves obtained with the $V_{2}$(BC), $V_{3}$(BM-E), and $V_{3}$(BM-EF) 
potentials are also very close to the observations. We notice that the $V_{2}$(BC) 
model introduced in Ref.~[\onlinecite{azizbc}] was constructed to reproduce the equation of state
calculated with the DFT-D2 method and that the good agreement displayed in Fig.~\ref{eos} is not a 
new result. Contrarily, the $V_{2}$, $V_{3}$(BM), and $V_{3}$(BM-F) potentials provide a poor 
description of the variation of the volume under pressure. In particular, we find that the $V_{3}$(BM) 
potential systematically underestimates $V$ at pressures equal or larger than $20$~GPa, in accordance 
with previous results reported by other authors.~\cite{chang01,herrero06} 
Meanwhile, the $V_{2}$ and $V_{3}$(BM-F) interaction models significantly overestimate the same 
quantity at pressures also close to or larger than $20$~GPa. In this latter case, we notice a 
surprising resemblance between the two calculated $P(V)$ curves.      

The main conclusion emerging from this part of our study is that the new $V_{3}$(BM-E) and 
$V_{3}$(BM-EF) three-body potentials reproduce very accurately the equation of state of solid 
helium up to a pressure of $\sim 60$~GPa (and possibly beyond). To the best of our knowledge, 
such a good agreement between theory and experiments has not been reported before for any  
known $V_{3}$ potential in solid $^{4}$He (see work~[\onlinecite{chang01}]).

\subsection{Kinetic energy}
\label{subsec:kinetic}
Our kinetic energy, $E_{\rm k}$, results are shown in Fig.~\ref{kinetice}.
In our DFT-D2 calculations, the kinetic energy was estimated within the 
quasiharmonic approximation through the expression
\begin{equation}
E_{\rm k}^{\rm qh} (V) = \frac{1}{N_{\rm q}} \sum_{\boldsymbol{q}s}
\frac{1}{2}\hbar\omega_{\boldsymbol{q}s}(V)~,
\label{eq:zpe}
\end{equation}
where $\omega_{\boldsymbol{q}s}$ are the vibrational phonon frequencies in the crystal
calculated at wave vector $\boldsymbol{q}$ and phonon branch $s$, which depend on the  
volume, and $N_{q}$ the total number of wave vectors used for integration within the 
first Brillouin zone (see Sec.~\ref{subsec:dft} and works~[\onlinecite{azizbc,cazorla13b}]). 
$E_{\rm k}^{\rm qh}$ usually is referred to as the ``zero-point energy'' (ZPE) and in many 
computational studies turns out to be crucial for predicting accurate solid-solid phase transitions.~\cite{shevlin12,cazorla08c,cazorla09} 
Regarding our DMC calculations, we computed first the \emph{exact} potential energy, $E_{\rm p}$, 
by means of the pure estimator technique~\cite{barnett91,casulleras95} and subsequently obtained the 
\emph{exact} kinetic energy by subtracting $E_{\rm p}$ to the corresponding total energy. In all the cases, 
spline interpolations were applied to the calculated data points in order to obtain smooth $P$-dependent 
energy curves (lines in Fig.~\ref{kinetice}). 

As is appreciated in the figure, the DFT-D2 results differ enormously from the rest 
of $E_{\rm k}$ series obtained with pairwise and three-body potentials in our DMC 
simulations. At the highest analysed pressure, for instance, the DFT-D2 kinetic energy is
a factor of two larger than the obtained DMC value. Given the lack of experimental data in the thermodynamic 
regime of interest, we can not rigorously conclude which type of calculation is providing the 
most realistic description. Nevertheless, we think that the DFT-D2 results are overestimating 
$E_{\rm k}$ severely because they have been obtained using the quasiharmonic approximation. 
In fact, it has been already demonstrated that the quasiharmonic approximation is not appropriate for studying 
crystals that behave much more classically than solid helium like, for instance, molecular hydrogen,~\cite{geneste12,mcmahon12,morales13} 
ammonia,~\cite{datchi06,pickard08} and some alkali metals.~\cite{errea11,feng15} It is worth noticing here 
that although the quasiharmonic DFT-D2 approach can produce equations of state that are in very good agreement 
with experiments (as it has been shown in Sec.~\ref{subsec:energy}), the accompanying ZPE corrections 
have a lot of margin for error since at high $P$ these are always several orders of magnitude smaller 
than the energy of the perfect crystal lattice. We shall comment again on this point in the next paragraph.  

\begin{figure}
\centerline
        {\includegraphics[width=1.0\linewidth]{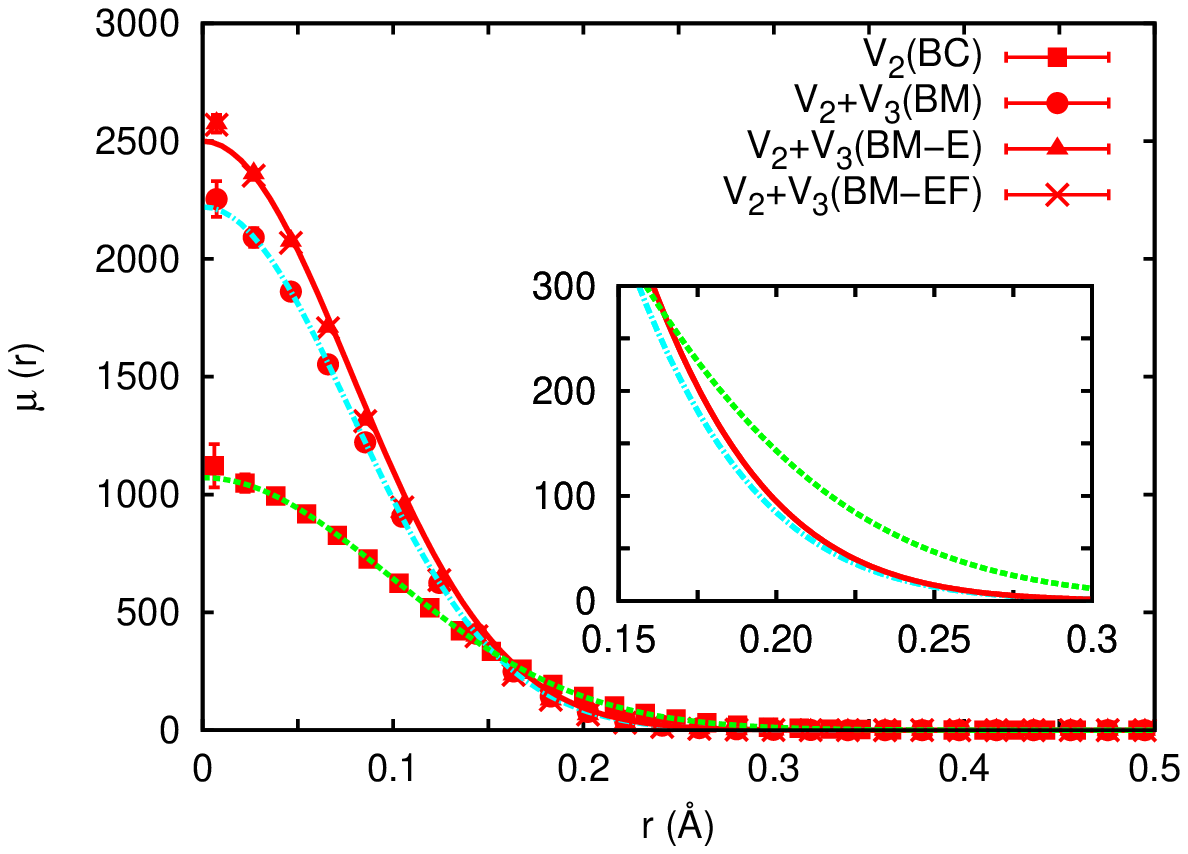}}
        {\includegraphics[width=1.0\linewidth]{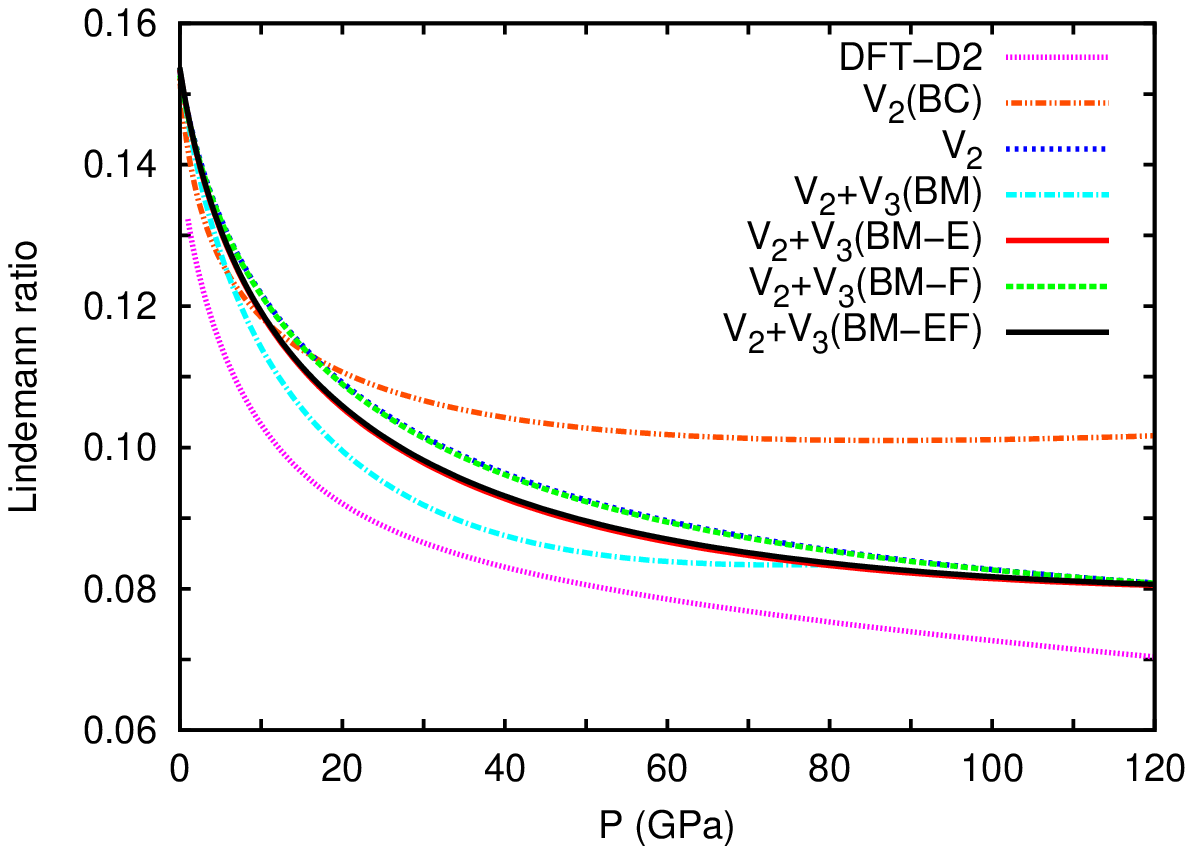}}
\caption{(\emph{Top})~Atomic density profile around the perfect lattice positions
         calculated with the DMC method considering different pairwise and
         three-body interaction models ($V = 3.0$~\AA$^{3}$/atom). Solid lines
         correspond to Gaussian curves fitted to the results. The corresponding
         tails are magnified in the inset in order to better appreciate the differences.
         (\emph{Bottom})~Lindeman ratio calculated in solid $^{4}$He with the DFT-D2 
         and DMC methods, expressed as a function of pressure.} 
\label{murlin}
\end{figure}

It is interesting to analyse the differences found between the (full quantum) DMC results obtained with 
different pairwise and three-body potential models. The $V_{2}$(BC) curve shows a plateau 
around $550$~K at pressures equal and beyond $\sim 80$~GPa. In a recent work,~\cite{azizbc} we identified 
such an infinitesimal variation in the kinetic energy with the presence of extreme quantum nuclear 
effects. However, calculations performed with the new set of three-body potentials introduced 
in this work bring new light into our previous interpretation of the $V_{2}$(BC) results. As 
is observed in Fig.~\ref{kinetice}, the $V_{3}$(BM-E), $V_{3}$(BM-F), and $V_{3}$(BM-EF) curves 
consistently display a small but steady increase in the kinetic energy under compression. At 
pressures below $\sim 15$~GPa the pairwise and three-body interaction models roughly provide 
equivalent $E_{\rm k}$ results however at $P = 160$~GPa the differences between 
them are as a large of $\sim 300$~K, with the $V_{3}$ potentials providing always the largest values. 
Several conclusions can be drawn from these results. First, although attenuated pairwise potentials 
based on exponential prefactors~\cite{moraldi12} can fairly reproduce experimental $P(V)$ data,~\cite{azizbc,omiyinka13} 
they are likely to introduce unwanted bias on the calculation of the kinetic energy. 
And second, the large $E_{\rm k}$ discrepancies observed between the DFT-D2 and $V_{3}$ results do not 
seem to be originated by the absence of four-, five- and so on many-body interactions in the DMC 
calculations. Actually, by comparing the energy curves obtained in the $V_{2}$ and $V_{2} + V_{3}$ 
cases one realizes that the effect of considering three-body interactions on $E_{\rm k}$ is rather small 
[only in the $V_{3}$(BM) case those effects are not negligible, although certainly minor]. Therefore, it 
is reasonable to expect similar trends when eventually one would add higher order many-body terms in the 
description of the atomic interactions. In regard to this last point, we notice that one of 
the main conclusions presented in work~[\onlinecite{azizbc}], namely that the quasiharmonic DFT approach 
exceedingly overestimates $E_{\rm k}$ in dense $^{4}$He, appears to be valid.

\subsection{Structural properties}
\label{subsec:elasticity}

An analysis of the atomic structure in solid $^{4}$He at high pressure will allow us to understand better the 
origins of the discrepancies found so far between the $V_{2}$(BC) and $V_{3}$ potentials. Figure~\ref{murlin} shows 
the atomic density profiles, $\mu (r)$, and Lindeman ratio, $\gamma$, calculated using the DMC method and several
atomic interaction models. The $\mu (r)$ results (see top panel) were obtained at volume $V = 3.0$~\AA$^{3}$/atom 
and subsequently were fitted to Gaussian functions (solid lines in the figure). As is observed there, the $V_{2}$(BC) 
curve is noticeably broader than all the others, and its value at the origin is about $50$~\% of that calculated with the 
$V_{3}$(BM) potential. Meanwhile, the $V_{3}$(BM-E) and $V_{3}$(BM-EF) profiles are practically indistinguishable 
and slightly higher near zero than the one obtained in the $V_{3}$(BM) case. Clearly, the $V_{2}$(BC) potential
produces a much larger atomic delocalization than the rest of interaction models, which is consistent with 
the kinetic energy results explained in the previous section. 

As for the Lindeman ratio $\gamma$ (see bottom panel in Fig.~\ref{murlin}), we have estimated the corresponding dependence
on pressure for each analysed potential. In the DFT-D2 case, $\gamma$ was computed within the quasiharmonic
approximation using the formula $9 \hbar^{2} / 8 m_{\rm He} E_{\rm k}^{\rm qh}$, see Eq.~(\ref{eq:zpe}) and 
works~[\onlinecite{cazorla08b,arms03}]. The results obtained in the $V_{2}$(BC) case are already known: a plateau 
around $0.10$ appears at pressures larger than $\sim 80$~GPa.~\cite{azizbc} However, all the other interaction models, 
including $V_{2}$ and $V_{3}$(BM), provide much smaller values of $\gamma$ at similar conditions. Moreover, the 
computed Lindeman ratio curves get depleted when compression is raised [with the exception of $V_{3}$(BM), 
in which $\gamma$ saturates around $0.08$ at pressures larger than $\sim 50$~GPa]. This latter trend is also 
observed in the DFT-D2 series, which systematically lies below the DMC predictions. 

The results presented in this section show that the $V_{2}$(BC) potential produces an unusually large delocalization 
of the atoms, which is at odds with the trends realised in the rest of cases. Such a huge particle dispersion effect 
is the responsible for the flat kinetic energy curve appearing in Fig.~\ref{kinetice}, which is likely to be an artifact 
deriving from the use of exponential attenuation factors at short distances.       

\begin{figure}
\centerline
        {\includegraphics[width=1.0\linewidth]{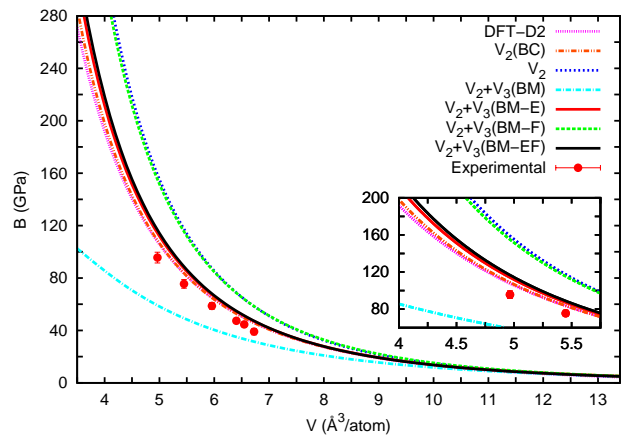}}
\caption{Calculated bulk modulus in $^{4}$He using the DFT-D2 and DMC
         methods and expressed as a function of volume. In the DMC case,
         different pairwise and three-body interaction models have been
         considered. Experimental data from work~[\onlinecite{zha04}]
         are shown for comparison. \emph{Inset}: The high-$P$ region in the
         $B(P)$ curves are magnified in order to appreciate better the
         differences.}
\label{bulkmodulus}
\end{figure}

\begin{figure}
\centerline
        {\includegraphics[width=1.0\linewidth]{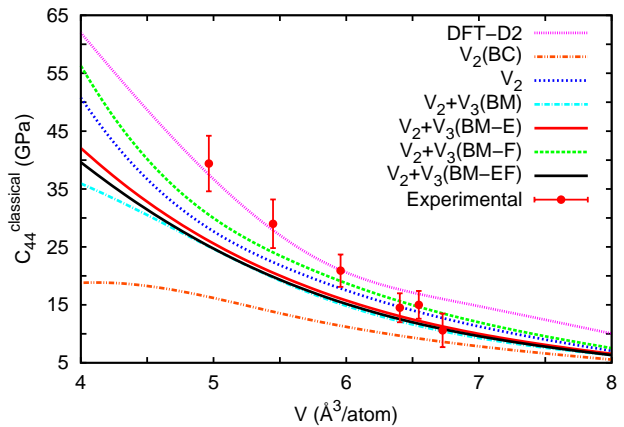}}
\caption{Calculated shear modulus in $^{4}$He using the DFT-D2 method and
         several force fields considering the atoms immobile in the perfect
         lattice positions. Experimental data are from work~[\onlinecite{zha04}].}
\label{c44}
\end{figure}

\subsection{Elastic properties}
\label{subsec:structure}

In Figs.~\ref{bulkmodulus} and~\ref{c44}, we show the bulk and shear modulii, $B$ and $C_{44}$ 
respectively, calculated in solid helium under pressure. The bulk modulus was directly 
obtained from the Birch-Murnaghan fits explained in Sec.~\ref{subsec:energy}, and in the 
$C_{44}$ case spline interpolations were applied to the calculated data points in order 
to obtain smooth $V$-dependent curves.   

Concerning the analysis of our $B(V)$ results, this is very much similar to the
conclusions presented for the equation of state in Sec.~\ref{subsec:energy}.  
Essentially, the DFT-D2, $V_{2}$(BC), $V_{3}$(BM-E), and $V_{3}$(BM-EF) curves are 
in good agreement with experiments whereas the $V_{2}$, $V_{3}$(BM-F), and $V_{3}$(BM) 
curves are not. In this latter case, both $V_{2}$ and $V_{3}$(BM-F) series are 
very similar and significantly overestimate the bulk modulus at small volumes. 
Likewise, the $V_{3}$(BM) potential provides unrealistically small values of $B(V)$
at large densities. 

\begin{table*}
\begin{center}
\label{tab:overview}
\begin{tabular}{c c c c c c}
\hline
\hline
$ $ & $ $ & $ $ & $ $ & $ $ & $ $ \\
$\quad \quad$ & $\quad P(V) \quad$ & $\quad B(V) \quad$ & $\quad C_{44}(V) \quad$ & $\quad {\rm E_{k}}/\gamma \quad$ & $\quad {\rm General~performance} \quad $ \\
$ $ & $ $ & $ $ & $ $ & $ $ & $ $ \\
\hline
$ $ & $ $ & $ $ & $ $ & $ $  & $ $ \\
$V_{2}$~[\onlinecite{aziz}] & $\times $ & $\times $ & $\surd/\times$ &  $\surd~(?) $ & ${\rm Not~satisfactory} $ \\
$ $ & $ $ & $ $ & $ $ & $ $  & $ $ \\
$V_{2}$(BC)~[\onlinecite{azizbc}] & $\surd $ & $\surd $ & $\times $ & $\times $ & ${\rm Not~satisfactory}$ \\
$ $ & $ $ & $ $ & $ $ & $ $  & $ $ \\
$V_{2}+V_{3}$(BM)~[\onlinecite{bm73}] & $\times $ & $\times $ & $\surd/\times $ & $\surd~(?) $ & ${\rm Not~satisfactory} $ \\
$ $ & $ $ & $ $ & $ $ & $ $ & $ $ \\
$V_{2}+V_{3}$(BM-E)       & $\surd $ & $\surd $ & $\surd/\times $ & $\surd~(?) $ & ${\rm Overall~good} $ \\
$ $ & $ $ & $ $ & $ $ & $ $ & $ $ \\
$V_{2}+V_{3}$(BM-F)       & $\times $ & $\times $ & $\surd/\times $ & $\surd~(?) $ & ${\rm Not~satisfactory} $ \\
$ $ & $ $ & $ $ & $ $ & $ $ & $ $ \\
$V_{2}+V_{3}$(BM-EF)      & $\surd $ & $\surd $ & $\surd/\times $ &  $\surd~(?) $ & ${\rm Overall~good} $ \\
$ $ & $ $ & $ $ & $ $ & $ $ & $ $ \\
\hline
\hline
\end{tabular}
\end{center}
\caption{Summary of the performance of the pairwise and three-body atomic interaction models analysed in this work
         in describing the energy, structural, and elastic properties of solid $^{4}$He at high pressure. Symbol
         $\surd$ ($\times$) indicates correct (incorrect) description of the considered quantity, whereas $\surd/\times$
         means quantitatively correct up to a certain pressure. Question mark ``$?$'' denotes a certain hesitation
         due to lack of experimental data in the high pressure regime of interest.}
\end{table*}

Let us now comment on the $C_{44} (V)$ results shown in Fig.~\ref{c44}. 
All the values have been obtained considering the atoms fixed on their perfect lattice positions, 
that is, totally neglecting likely quantum nuclear effects (hence the employed subscript). 
This is done for the sake of comparison since it is technically difficult to account for quantun 
nuclear effects in the DFT-D2 calculations in an exact manner, that is, to go beyond the quasiharmonic 
approximation. Nevertheless, later on this section we will show that according to our DMC simulations 
quantum nuclear effects become secondary on $C_{44}$ at high pressure.
As is observed in the figure, the DFT-D2 curve is in overall good agreement with the ambient temperature 
measurements performed by Zha and collaborators.~\cite{zha04} Again, these findings justify our choice
of the benchmark data for the modeling of three-body interactions. Regarding the performance of the 
original and new three-body BM potentials, we find that in general they reproduce quite satisfactorily the 
experimental data obtained at volumes larger than $\sim 5.5$~\AA$^{3}$/atom (i.e., $P \le 25$~GPa). 
This is especially true in the $V_{3}$(BM-F) case where, as expected (see Sec.~\ref{sec:potentialfit}), 
the calculated shear modulii follow closely those obtained with the DFT-D2 method. However, at volumes smaller 
than $\sim 5.5$~\AA$^{3}$/atom (i.e., $P \ge 25$~GPa) we find that the differences between the BM 
curves (including the $V_{3}$BM-F case), on one side, and the DFT-D2 results and experiments, on the other, become 
increasingly larger. We recall that the $V_{3}$(BM-E) and $V_{3}$(BM-EF) potentials provide a very good description 
of the equation of state and bulk modulus, whereas the $V_{3}$(BM-F) potential does not. This appreciation 
let us to conclude that is very difficult to provide simultaneously a good account of the energy and 
elastic properties in solid helium by using an effective three-body approach. Higher order many-body contributions 
in the description of the atomic interactions probably are necessary in order to attain an overall correct 
description of solid helium at high pressure. As for the pairwise potentials, $V_{2}$ performs very similarly 
to the $V_{3}$(BM-F) model, as we have also noted in the total energy (see Sec.~\ref{subsec:energy}) and 
bulk modulus cases. The $V_{2}$(BC) model, however, remarkably fails in reproducing the variation of the shear modulus
under pressure. Moreover, it predicts the occurrence of unrealistic mechanical instabilities 
(i.e., $d C_{44}/ d V \approx 0$)~\cite{sinko02,grimvall12} at small volumes. Therefore, the use of the $V_{2}$(BC) 
potential is strongly not recommended for the simulation of solid helium at high pressure.      
  
In order to quantify the importance of quantum nuclear effects on the calculation of the shear modulus, we 
carried out additional quantum DMC calculations (see Sec.~\ref{subsec:dmc} and works~[\onlinecite{cazorla12,cazorla13,cazorla12b}] 
for details). To our surprise, we found that the quantum and classical shear modulii results are very similar. 
For instance, in the $V_{3}$(BM-F) case the $C_{44}^{\rm classical} - C_{44}^{\rm quantum}$ difference
(where subscript ``quantum'' means calculated with the DMC method) amounts only to $2$~GPa at $P \sim 50$~GPa. Similar 
results were obtained also in the rest of $V_{2}$ and $V_{3}$ cases. We note that the sign of the differences is always 
positive, thus the inclusion of quantum nuclear effects tends to lower the classical $C_{44}$ values, although in a 
small fraction (i.e., $\simeq 5$~\%). This last finding appears to be consistent with conclusions presented in 
a recent quantum Monte Carlo study by Borda \emph{et al.},~\cite{borda14} in which the ideal shear strength 
on the basal plane of hcp $^{4}$He was found to behave analogously than in classical solids.

\section{Conclusions}
\label{sec:conclusions}

In Table~III we summarise the performance of the analysed pairwise and three-body potentials in 
describing the energy, elastic and structural properties of solid $^{4}$He at high pressure. 
A number of tips can be drawn from our results. First of all, the use of pairwise potentials 
in general is not recommended. These either fail to reproduce the equation 
of state and bulk modulus, i.e., $V_{2}$, or the kinetic energy, and structural and elastic 
features, i.e., $V_{2}$(BC), in highly compressed quantum crystals. In this context, we urge 
to employ more versatile many-body interaction models. This is the case, for instance, of the 
new three-body BM potentials introduced in this work, which represent an improvement with 
respect to previously reported similar models. Overall, we recommend to consider the $V_{3}$(BM-E)
and $V_{3}$(BM-EF) parametrizations in prospective simulation studies because they provide the most 
satisfactory general description of dense solid $^{4}$He. Indeed, those interaction models can be safely 
employed, for instance, in atomistic high-$P$ high-$T$ simulations (either classical or quantum), 
which are of relevance to planetary sciences. Nevertheless, we must note that it remains a challenge 
to attain a precise description of elasticity at high pressure by using effective three-body potentials, 
thus in this latter case consideration of higher order many-body terms appears to be necessary.                   

Importantly, we have shown that the addition of three-body forces corrects for the artificially 
large atomic delocalization found with modified pairwise potentials based on exponential 
attenuation factors. Nevertheless, given the lack of structural and kinetic energy measurements performed 
at high pressure, we have not been able to quantify the accuracy of our $\gamma$ and $E_{\rm k}$  
DMC results obtained with the $V_{3}$(BM-E) and $V_{3}$(BM-EF) potential models. In this regard, 
advanced computational studies in which both the nuclear and electronic degrees of freedom in the
crystal were to be treated at the quantum level are highly desirable.

\begin{acknowledgments}
This research was supported under the Australian Research Council's Future Fellowship funding scheme
(projects number RG134363 and RG151175), and MICINN-Spain (Grants No. MAT2010-18113, CSD2007-00041, and 
FIS2014-56257-C2-1-P). 
\end{acknowledgments}


\begin{thebibliography}{30}
\bibitem{kalos81} M. H. Kalos, M. A. Lee, and P. A. Whitlock, Phys. Rev. B \textbf{24}, 115 (1981). 
\bibitem{boronat94} J. Boronat and J. Casulleras, Phys. Rev. B \textbf{49}, 8920 (1994).
\bibitem{aziz} R. A. Aziz, F. R. W. McCourt, and C. C. K. Wong, Mol. Phys. \textbf{61}, 1487 (1987).		
\bibitem{cazorla08} C. Cazorla and J. Boronat, J. Phys.: Condens. Matt. \textbf{20}, 015223 (2008). 
\bibitem{moraldi12} M. Moraldi, J. Low Temp. Phys. \textbf{168}, 275 (2012).
\bibitem{azizbc} C. Cazorla and J. Boronat, Phys. Rev. B \textbf{91}, 024103 (2015).
\bibitem{omiyinka13} T. Omiyinka and M. Boninsegni, Phys. Rev. B \textbf{88}, 024112 (2013).
\bibitem{wallace72} D. C. Wallace, \textit{Thermodynaimcs of Crystals} (Wiley, New York, 1972).
\bibitem{pechenik08} E. Pechenik, I. Kelson, and G. Makov, Phys. Rev. B \textbf{78}, 134109 (2008).
\bibitem{freiman09} Y. A. Freiman, S. M. Tretyak, A. Grechnev, A. F. Goncharov, J. S. Tse, D. Errandonea, 
                    H.-K. Mao, and R. J. Hemley, Phys. Rev. B \textbf{80}, 094112 (2009). 
\bibitem{zha04} C.-S. Zha, H.-K. Mao, and R. J. Hemley, Phys. Rev. B \textbf{70}, 174107 (2004).
\bibitem{bm73} L. W. Bruch and I. J. McGee, J. Chem. Phys. \textbf{59}, 409 (1973).
\bibitem{cohen96} M. J. Cohen and J. N. Murrel, Chem. Phys. Lett. \textbf{260}, 371 (1996).
\bibitem{loubeyre86} P. Loubeyre, Phys. Rev. Lett. \textbf{58}, 1857 (1986).
\bibitem{chang01} S.-Y. Chang and M. Boninsegni, J. Chem. Phys. \textbf{115}, 2629 (2001).
\bibitem{herrero06} C. P. Herrero, J. Phys.:Condens. Matt. \textbf{18}, 3469 (2006).
\bibitem{tian06} C.-L. Tian, F.-S. Liu, F.-Q. Jing, and L.-C. Cai, 
                 J. Phys.:Condens. Matt. \textbf{18}, 8103 (2006). 
\bibitem{grimme06} S. Grimme, J. Comp. Chem. \textbf{27}, 1787 (2006).
\bibitem{pbe96} J. P. Perdew, K. Burke, and M. Ernzerhof,
                Phys. Rev. Lett. \textbf{77}, 3865 (1996).
\bibitem{vasp} G. Kresse and J. F\"urthmuller, Phys. Rev. B \textbf{54}, 11169 (1996).
\bibitem{cazorla15} C. Cazorla, Coord. Chem. Rev. \textbf{300}, 142 (2015).
\bibitem{blochl94} P. E. Bl\"{o}chl, Phys. Rev. B {\bf 50}, 17953 (1994).
\bibitem{kresse99} G. Kresse and D. Joubert, Phys. Rev. B {\bf 59}, 1758 (1999).
\bibitem{cazorla07} C. Cazorla, M. J. Gillan, S. Taioli, and D. Alf\`e,
                    J. Chem. Phys. {\bf 126}, 194502 (2007).
\bibitem{taioli07} S. Taioli, M. J. Gillan, C. Cazorla, and D. Alf\`e,
                   Phys. Rev. B {\bf 75}, 214103 (2007).
\bibitem{cazorla13b} C. Cazorla and J. ${\rm \acute{I}}$${\rm \tilde{n}}$iguez,
                    Phys. Rev. B \textbf{88}, 214430 (2013).
\bibitem{shevlin12} S. A. Shevlin, C. Cazorla, and Z. X. Guo,
                    J. Phys. Chem. C \textbf{116}, 13488 (2012).
\bibitem{cazorla08c} C. Cazorla, D. Alf\`e, and M. J. Gillan,
                    Phys. Rev. Lett. {\bf 101}, 049601 (2008).
\bibitem{alfe09} D. Alf\`e, Comp. Phys. Commun. \textbf{180}, 2622 (2009).
\bibitem{cazorla09b} C. Cazorla, G. Astrakharchick, J. Casulleras, and J. Boronat,
                     New Journal of Phys. \textbf{11}, 013047 (2009).
\bibitem{cazorla08b} C. Cazorla and J. Boronat, Phys. Rev. B \textbf{77}, 024310 (2008).
\bibitem{cazorla10}  C. Cazorla, G. Astrakharchick, J. Casulleras, and J. Boronat,
                     J. Phys.: Condens. Matter \textbf{22}, 165402 (2010).
\bibitem{gordillo11} M. C. Gordillo, C. Cazorla, and J. Boronat,
                     Phys. Rev. B \textbf{83}, 121406(R) (2011).
\bibitem{chin90} S. A. Chin, Phys. Rev. A \textbf{42}, 6991 (1990).
\bibitem{cazorla12} C. Cazorla, Y. Lutsyshyn, and J. Boronat,
                    Phys. Rev. B \textbf{85}, 024101 (2012).
\bibitem{cazorla13} C. Cazorla, Y. Lutsyshyn, and J. Boronat,
                    Phys. Rev. B \textbf{87}, 214522 (2013).
\bibitem{cazorla12b} R. Rota, Y. Lutsyshyn, C. Cazorla, and J. Boronat,
                     J. Low Temp. Phys. \textbf{168}, 150 (2012).
\bibitem{ercolessi94} F. Ercolessi and J. B. Adams, Europhys. Lett. \textbf{26}, 583 (1994).
\bibitem{marco12} J. Sala, E. Gu\`ardia, J. Mart\'i, D. Sp\aa ngberg, and M. Masia, 
                  J. Chem. Phys. \textbf{136}, 054103 (2012). 
\bibitem{marco14} M. Masia, E. Gu\`ardia, and P. Nicolini, Int. J. Quantum. Chem. \textbf{114}, 1036 (2014).
\bibitem{nocedal06} J. Nocedal and S. J. Wright, \textit{Numerical Optimization} 
                    (Springer, Berlin, 2006).
\bibitem{loubeyre93} P. Loubeyre, R. LeToullec, J. P. Pinceaux, H. K. Mao, J. Hu, and R. J. Hemley,
                     Phys. Rev. Lett. \textbf{71}, 2272 (1993).
\bibitem{birch78} F. Birch, J. Geophys. Res. \textbf{83}, 1257 (1978).
\bibitem{cazorla09} C. Cazorla, D. Errandonea, and E. Sola, Phys. Rev. B \textbf{80}, 064105 (2009).
\bibitem{barnett91} R. Barnett, P. Reynolds, and W. A. Lester Jr.,
                    J. Comput. Phys. \textbf{96}, 258 (1991).
\bibitem{casulleras95} J. Casulleras and J. Boronat,
                       Phys. Rev. B \textbf{52}, 3654 (1995).
\bibitem{geneste12} G. Geneste, M. Torrent, F. Bottin, and P. Loubeyre,
                    Phys. Rev. Lett. \textbf{109}, 155303 (2012).
\bibitem{mcmahon12} J. M. McMahon, M. A. Morales, C. Pierleoni, and D. M. Ceperley,
                    Rev. Mod. Phys. \textbf{84}, 1607 (2012).
\bibitem{morales13} M. A. Morales, J. M. McMahon, C. Pierleoni, and D. M. Ceperley,
                    Phys. Rev. B \textbf{87}, 184107 (2013).
\bibitem{datchi06} F. Datchi, S. Ninet, M. Gauthier, A. M. Saitta, B. Canny, and F. Decremps, 
                   Phys. Rev. B \textbf{73}, 174111 (2006).
\bibitem{pickard08} C. J. Pickard and R. J. Needs, Nature Materials \textbf{7}, 775 (2008).
\bibitem{errea11} I. Errea, B. Rousseau, and A. Bergara,
                  Phys. Rev. Lett. \textbf{106}, 165501 (2011).
\bibitem{feng15} Y. Feng, J. Chen, D. Alf\`e, X-Z. Li, and E. Wang, J. Chem. Phys. \textbf{142}, 064506 (2015).
\bibitem{arms03} D. A. Arms, R. S. Shah, and R. O. Simmons,
                 Phys. Rev. B \textbf{67}, 094303 (2003).
\bibitem{sinko02} G. V. Sin'ko and N. A. Smirnov,
                  J. Phys.: Condens. Matter \textbf{14}, 6989 (2002).
\bibitem{grimvall12} G. Grimvall \emph{et al.}, Rev. Mod. Phys. \textbf{84}, 945 (2012).
\bibitem{borda14} E. J. L. Borda, W. Cai, and M. de Koning, Phys. Rev. Lett. \textbf{112}, 155303 (2014).
\end{thebibliography}
\end{document}